\documentclass[10pt, conference]{IEEEtran}
\IEEEoverridecommandlockouts
\usepackage{cite}
\usepackage{amsmath,amssymb,amsfonts}
\usepackage{algorithmic}
\usepackage{graphicx}
\usepackage{textcomp}
\usepackage{xcolor}
\usepackage{url}
\usepackage{siunitx}
\def\BibTeX{{\rm B\kern-.05em{\sc i\kern-.025em b}\kern-.08em
    T\kern-.1667em\lower.7ex\hbox{E}\kern-.125emX}}

\usepackage[acronym,shortcuts]{glossaries}
\newacronym{AMD}{AMD}{age-related macular degeneration}
\newacronym{MSE}{MSE}{mean squared error}
\newacronym{RP}{RP}{retinitis pigmentosa}
\glsdisablehyper

\begin{document}

\title{\LARGE \bf Predicting the Temporal Dynamics of Prosthetic Vision

\author{Yuchen Hou$^{1,*}$, Laya Pullela$^{1,*}$, Jiaxin Su$^{2,4}$, Sriya Aluru$^{1}$, Shivani Sista$^{1}$, Xiankun Lu$^{3}$, and Michael Beyeler$^{1,4}$}
\thanks{$^{*}$These authors contributed equally to this work.}
\thanks{$^{1}$YH, LP, SA, SS, and MB are with the Department of Computer Science, University of California, Santa Barbara, CA 93106, USA.}
\thanks{$^{2}$JS is with the Department of Statistics and Applied Probability, University of California, Santa Barbara, CA 93106, USA.}
\thanks{$^{3}$XL is with the Department of Physics, University of California, Santa Barbara, CA 93106, USA.}
\thanks{$^{4}$JS and MB are with the Department of Psychological \& Brain Sciences, University of California, Santa Barbara, CA 93106, USA.}
\thanks{The authors confirm contribution to the paper as follows. Study conception and design: MB; model development: YH, LP, SA, MB; data processing and analysis: YH, LP, JS, SA, SS, XL, MB; manuscript writing: YH, LP, MB. All authors reviewed the results and approved the final version of the manuscript.}
\thanks{This work was partially supported by the National Institutes of Health (NIH R00 EY-029329 to MB). We thank Jacob Granley and Justin Kasowski for their valuable contributions to an earlier version of this work.}
}


\maketitle

\begin{abstract}

Retinal implants are a promising treatment option for degenerative retinal disease. While numerous models have been developed to simulate the appearance of elicited visual percepts (``phosphenes''), these models often either focus solely on spatial characteristics or inadequately capture the complex temporal dynamics observed in clinical trials, which vary heavily across implant technologies, subjects, and stimulus conditions. Here we introduce two computational models designed to accurately predict phosphene fading and persistence under varying stimulus conditions, cross-validated on behavioral data reported by nine users of the Argus II Retinal Prosthesis System. Both models segment the time course of phosphene perception into discrete intervals, decomposing phosphene fading and persistence into either sinusoidal or exponential components. Our spectral model demonstrates state-of-the-art predictions of phosphene intensity over time (\textit{r}~=~0.7 across all participants). 
Overall, this study lays the groundwork for enhancing prosthetic vision by improving our understanding of phosphene temporal dynamics.

\end{abstract}


\section{Introduction}

Retinal prostheses offer a technological solution to the vision loss caused by photoreceptor degeneration \cite{weiland_electrical_2016}, as seen in conditions like retinitis pigmentosa (RP) and dry age-related macular degeneration (geographic atrophy).
These devices electrically stimulate the retina to evoke the perception of phosphenes (localized spots of light) \cite{erickson-davis_what_2021}.
Although the \emph{spatial} appearance of these phosphenes is well understood  \cite{beyeler_model_2019,granley_computational_2021}, the \emph{temporal} dynamics remain a challenge.
Implant technologies that do not incorporate eye movements tend to generate phosphenes that either fade in less than a second or persist for several seconds after stimulus offset \cite{perez_fornos_temporal_2012}.
The variability in the duration of phosphenes is thought to be influenced by retinal ganglion cell (RGC) desensitization and adaptation to static stimuli \cite{freeman_multiple_2011,li_retinal_2022,johnston_rapid_2024}, thereby complicating the continuous perception needed for effective vision in everyday activities.

Previous studies have investigated the temporal properties of phosphene perception, ranging from clinical observations to computational simulations aimed at elucidating the physiological and neuroanatomical bases of these dynamics. 
Perez-Fornos and colleagues \cite{perez_fornos_temporal_2012} investigated the temporal properties of phosphene perception for users of the Argus II Retinal Prosthesis System (Vivani Medical; formerly Second Sight Medical Products, Inc.).
Participants were asked to indicate the perceived phosphene brightness over time using a joystick, while being stimulated with pulse trains of different rates and duration.
Only one out of nine subjects reported a stable percept appearing at stimulus onset and lasting until stimulus offset; the others reported phosphene brightness to change more or less rapidly depending on stimulus conditions, sometimes also changing shape and color over time.
Computational efforts have been made to model phosphene temporal sensitivity and predict brightness responses, often convolving the stimulus with linear kernels and nonlinear activation functions \cite{horsager_predicting_2009} to explain phenomena such as persistence and fading \cite{avraham_retinal_2021,grinten_biologically_2022}.
However, these models have often been limited by their reliance on simplified assumptions or the absence of validation against clinically reported, real-world data.

To address this gap, our paper introduces two computational models designed to more accurately predict the complex temporal dynamics of phosphene perception. 
The first model applies traditional exponential and sine decay functions to forecast phosphene brightness based on specific stimulus parameters, whereas the second leverages the Fast Fourier Transform (FFT) to better model the interplay of persistence and fading, similar to work from the cochlear implant community \cite{hu_model_2023}.
These models stand out as the first to be directly validated against the complex temporal features of real subjects' data, which help inform future retinal stimulation strategies.



\begin{figure*}[th!]
    \centering
    \includegraphics[width=\linewidth]{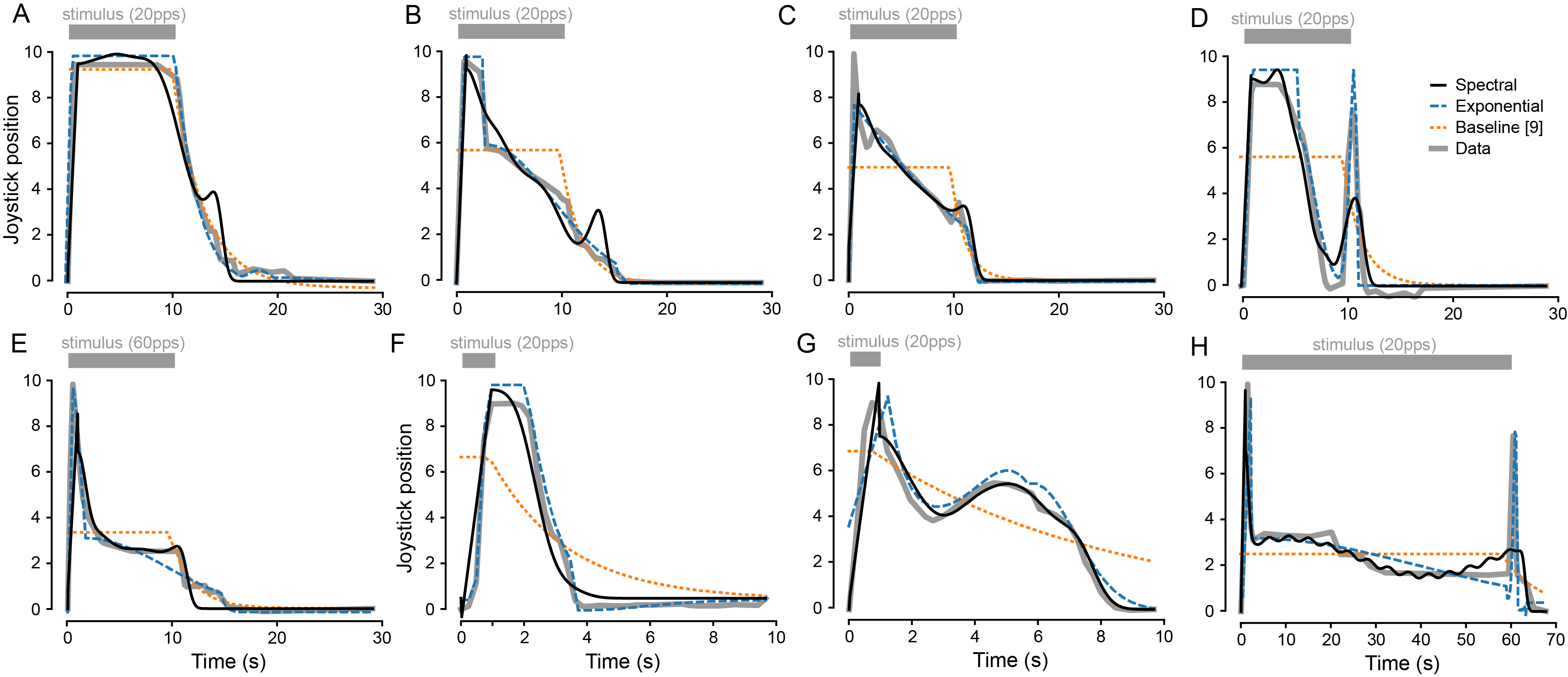}
    \caption{Model fits for representative time courses, each elicited by distinct electrical stimuli (see Table II for details). Subjects' perceived brightness levels (gray; indicated by joystick position \cite{perez_fornos_temporal_2012}) are depicted alongside predictions from three models: spectral (black), exponential (blue), and baseline \cite{avraham_retinal_2021} (orange). }
    \label{fig:results-model-fits}
\end{figure*}

\section{Methods}

\subsection{Spectral Model}
Phosphenes elicited by retinal implants typically give rise to a brisk increase in brightness upon stimulus onset, followed by a decay phase at various rates. 
The spectral model conceptualizes the brightness time course of phosphenes as a piecewise function, segmented by time intervals $t_1 \dots t_3$:



\begin{equation}
I[t]= \begin{cases} 
k_1 (t-t_1) & \text{if } t \in [0, t_1) \\
\frac{I[t_2] - k_1}{t_2-t_1} (t-t_1) + k_1 & \text{if } t \in [t_1, t_2) \\
\mathcal{F}^{-1}(X_m) + k_2 & \text{if } t \in [t_2, t_3) \\
0 & \text{if } t \in [t_3, \infty)
\end{cases}
\end{equation}

The model assumes a linear rise (with slope $k_1=10$) until maximum brightness is reached at $t_1=\SI{1}{\second}$, after which brightness decay is modeled as a truncated Fourier series. $X_m$ is the filtered discrete fourier transform (DFT) of $f[t_1:t_3]$ retaining only the top $m$ spectral components with the highest magnitudes; we found $m \in [2,4]$ was sufficient. $\text{ }\mathcal{F}^{-1}(X_m)$ is the inverse DFT of $X_m$, and $k_2$ represents the bias term (vertical shift). To connect the onset rise with the decay curve, we find $t_2$, the first local maxima of the decay curve, and make a shallow linear connector piece between time steps $[t_1,t_2)$. We optimize $t_3$ to force the brightness to zero at the time step where the percept essentially flatlines.
The open parameters of this model are $k_2$, $t_3$, and $m$.

\subsection{Exponential Model}

The exponential model separates the predicted brightness time course into six periods separated by time steps $t_1, \ldots, t_5$:

\begin{equation}
    I[t]= \begin{cases} 
0 & \text{if } t \in [0, t_\text{1}) \\ 
k_\text{1}  (t-t_1) & \text{if } t \in [t_1, t_2 ) \\ 
I[t-1]  \exp \big(-k_2 (t-t_2) f^{\frac{1}{4}} \big) & \text{if } t \in [t_2, t_3) \\ 
I[t-1]  \exp \big(-k_3 (t-t_3) \big) & \text{if } t \in [t_3, t_4) \\ 
 I[t-1]  k_4 \sin \big(k_5 (t-t_4) \big) & \text{if } t \in [t_4, t_5) \\ 
 I[t-1]  \exp \big(-k_3 (t-t_5) \big). & \text{if } t \in [t_5, \infty)  \, \\ 
\end{cases} 
\end{equation}

The model assumes a linear rise (with slope $k_1$) until maximum brightness is reached at $t_2$, after which brightness decays with exponential decay time constants $k_2 f^{\frac{1}{4}}$ and $k_3$, where $f$ is the stimulus frequency.
At time $t_4$, a second rise in brightness may appear (sometimes seen after stimulus offset; e.g., Fig.~\ref{fig:results-model-fits}D and Fig.~\ref{fig:results-model-fits}H), which is modeled as a sine wave whose magnitude and frequency are controlled by $k_4$ and $k_5$, respectively.
After that ($t \ge t_5$), brightness continues to decay at rate $k_3$.
All scaling parameters ($k_1, \ldots, k_5$) and time points ($t_1, \ldots, t_5$) were learned (i.e., ten open parameters).

\begin{table*}[t!]
    \centering
    \renewcommand{\arraystretch}{1.25}
\caption{The performance of the baseline, the exponential model, and the Spectral model across different subjects evaluated by the mean squared error (MSE) and Pearson's correlation coefficient (\textit{r}). The best-performing numbers for each subject are highlighted in bold.}
    \label{tab:results-per-subject}
\begin{tabular}{lrrlrrlrr}
\hline\hline
Subject & \multicolumn{2}{c}{Baseline} &  & \multicolumn{2}{c}{Exponential Model} &  & \multicolumn{2}{c}{Spectral Model ($m=2$)} \\
\cline{2-3} \cline{5-6} \cline{8-9}
 & MSE $\downarrow$ & \textit{r} $\uparrow$ &  & MSE $\downarrow$ & \textit{r} $\uparrow$ &  & MSE $\downarrow$ & \textit{r} $\uparrow$ \\
 \hline
1 & 8.342 $\pm$ 6.132 & 0.740 $\pm$ 0.215 &  & 8.134 $\pm$ 10.200 & 0.776 $\pm$ 0.204 &  & 
\textbf{3.644 $\pm$ 3.429} & \textbf{0.880 $\pm$ 0.031} \\
2 & 3.796 $\pm$ 1.397 & -0.124 $\pm$ 0.159 &  & \textbf{3.564 $\pm$ 1.309} & 0.119 $\pm$ 0.188 &  & 
4.232 $\pm$ 1.751 & \textbf{0.159 $\pm$ 0.378} \\
3 & 4.143 $\pm$ 2.729 & 0.762 $\pm$ 0.091 &  & 7.523 $\pm$ 4.769 & 0.672 $\pm$ 0.106 & & \textbf{4.099 $\pm$ 2.930} & \textbf{0.807 $\pm$ 0.084} \\
4 & 11.802 $\pm$ 5.182 & 0.402 $\pm$ 0.362 &  & \textbf{4.646 $\pm$ 3.736} & \textbf{0.731 $\pm$ 0.278} &  & 7.271 $\pm$ 2.703 & 0.596 $\pm$ 0.284 \\

5 & 10.130 $\pm$ 7.196 & 0.520 $\pm$ 0.350 &  & 4.557 $\pm$ 2.877 & 0.831 $\pm$ 0.103 &  & \textbf{4.470 $\pm$ 3.058} & \textbf{0.843 $\pm$ 0.080}\\


6 & 9.287 $\pm$ 8.478 & 0.789 $\pm$ 0.160 &  & \textbf{4.333 $\pm$ 2.746} & 0.717 $\pm$ 0.398 & & 8.782 $\pm$ 3.047 & \textbf{0.800 $\pm$ 0.128} \\

7 & 3.659 $\pm$ 1.014 & 0.193 $\pm$ 0.247 &  & 5.080 $\pm$ 3.523 & 0.271 $\pm$ 0.277 & & \textbf{3.066 $\pm$ 1.260} & \textbf{0.465 $\pm$ 0.220}\\

8 & 4.164 $\pm$ 3.280 & 0.752 $\pm$ 0.213 &  & 4.131 $\pm$ 3.331 & \textbf{0.847 $\pm$ 0.153} &  &\textbf{ 3.630 $\pm$ 1.128} & 0.827 $\pm$ 0.050 \\

9 & 1.700 $\pm$ 0.681 & 0.696 $\pm$ 0.209 &  & 2.056 $\pm$ 1.304 & 0.730 $\pm$ 0.163 & & \textbf{1.146 $\pm$ 0.531} & \textbf{0.861 $\pm$ 0.064}\\
\hline
Average & 6.336 $\pm$ 5.882 & 0.526 $\pm$ 0.381 &  & 4.892 $\pm$ 4.851 & 0.633 $\pm$ 0.332 &  &\textbf{4.482 $\pm$ 3.370} & \textbf{0.693 $\pm$ 0.312}\\

\hline \hline
\end{tabular}
\vspace{7mm}

\end{table*}
\begin{table*}[bt!]
    \centering
    \renewcommand{\arraystretch}{1.25}
    \caption{The performance of the Baseline, the Exponential model, and the Spectral model across different stimulus conditions evaluated by MSE and correlation (\textit{r}). The best-performing numbers for each stimulus condition are highlighted in bold.}
    \label{tab:results-per-stimulus}
    \begin{tabular}{llllllllllll}
    \hline\hline
    & \multicolumn{2}{c}{Stimulus Condition} &  & \multicolumn{2}{c}{Baseline} &  & \multicolumn{2}{c}{Exponential Model} &  & \multicolumn{2}{c}{Spectral Model ($m=2$)} \\
    \cline{2-3} \cline{5-6} \cline{8-9} \cline{11-12}
     & Freq (pps) & Stim Dur (s) &  & \multicolumn{1}{r}{MSE} $\downarrow$ & \multicolumn{1}{r}{\textit{r}} $\uparrow$ &  & \multicolumn{1}{r}{MSE} $\downarrow$ & \multicolumn{1}{r}{\textit{r}} $\uparrow$ &  & \multicolumn{1}{r}{MSE} $\downarrow$ & \multicolumn{1}{r}{\textit{r}} $\uparrow$ \\
     \hline 
    1 & 20 & 10 &  & 6.001 $\pm$ 4.648 & 0.656 $\pm$ 0.390 &  & \textbf{5.744 $\pm$ 5.621} & \textbf{0.686 $\pm$ 0.433} & & 5.983 $\pm$ 3.572 & 0.683 $\pm$ 0.256 \\ 
    
    2 & 5 & 10 &  & 6.960 $\pm$ 7.175 & 0.688 $\pm$ 0.264 &  & 6.327 $\pm$ 7.492 & 0.757 $\pm$ 0.255 & & \textbf{6.254 $\pm$ 4.807} & \textbf{0.760 $\pm$ 0.168}\\ 
    
    3 & 60 & 10 &  & 7.330 $\pm$ 5.833 & 0.541 $\pm$ 0.399 &  & \textbf{6.828 $\pm$ 5.626} & 0.566 $\pm$ 0.372 & & 7.057 $\pm$ 3.725 & \textbf{0.567 $\pm$ 0.229}\\
    
    4 & 20 & 1 &  & 7.549 $\pm$ 4.472 & 0.585 $\pm$ 0.212 &  & 7.270 $\pm$ 4.455 & \textbf{0.625 $\pm$ 0.315} & & \textbf{7.241 $\pm$ 3.725} & 0.612 $\pm$ 0.138 \\
    
    5 & 20 & 60 &  & 8.084 $\pm$ 4.899 & 0.455 $\pm$ 0.300 &  & 7.924 $\pm$ 4.724 & 0.512 $\pm$ 0.303 & & \textbf{7.548 $\pm$ 2.281} & \textbf{0.547 $\pm$ 0.143}\\
    
     \hline 
    \multicolumn{3}{l}{Average} &  & 7.185 $\pm$ 5.541 & 0.585 $\pm$ 0.332 &  & 6.818 $\pm$ 5.733 & 0.629 $\pm$ 0.352 & &\textbf{6.817 $\pm$ 5.631} & \textbf{0.634 $\pm$ 0.305}\\
    
    \hline \hline
    \end{tabular}
\end{table*}

\subsection{Baseline Model}
 
Avraham \emph{et al.}~\cite{avraham_retinal_2021} used a pair of exponential terms to model i) phosphene persistence (lasting $t_\text{PER}$ seconds, approached with exponential time constant $\tau_\text{PER}$):
\begin{equation}
{I_{\text{PER}}(t) = I_0 \exp\left(\frac{-t}{\tau_{\text{PER}}}\right) \small{\text{, with }} \tau_{\text{PER}} = \frac{t_\text{PER}}{\ln(I_0) - \ln(I_e)} }
\end{equation}
and ii) phosphene fading (starting at $t_\text{PFO}$ and lasting $t_\text{PFD}$ seconds, with time constant $\tau_\text{PDF}$):

\begin{equation}
{\begin{aligned}
I_{\text{PFD}}(t) &= I_0 \exp\left(-\frac{t-t_\text{PFO}}{\tau_{\text{PFD}}}\right), \\
\small{\text{with }} \tau_{\text{PFD}} &= \frac{t_\text{PFD}}{\ln(I_0) - \ln(I_e)}
\end{aligned}
}
\end{equation}

where $I_0$ is the initial normalized phosphene intensity and $I_e$ is the intensity of the background (here, we chose $2\%$ of $I_0$).
Any phosphene time course could thus be modeled by a combination of persistence and fading periods. We also introduced a scaling factor $k$ to allow for arbitrary brightness values.
To avoid flickering, phosphene intensity was resampled at a \SI{0.25}{\second} timestep.
Parameters $t_\text{PER}$, $t_\text{PFO}$, $t_\text{PFD}$ and $k$ were learned (i.e., the model has four open parameters).


\subsection{Dataset}
All models were fit to time-series data collected on nine Argus II users across five stimulus conditions \cite{perez_fornos_temporal_2012}.
During each trial, four neighboring electrodes (quad) were simultaneously stimulated, and subjects were asked to describe their current perceived brightness with a joystick.
Biphasic pulse trains varied in frequency (5, 20, or 60 pps) and stimulus duration (1, 10, or \SI{60}{\second}), and amplitudes were set at the upper comfortable level
\cite{perez_fornos_temporal_2012}.
Time courses were manually extracted by sampling points using the WebPlotDigitizer software and re-sampled at a \SI{0.25}{\second} timestep.
This dataset is available as part of the \emph{pulse2percept} Python package \cite{beyeler_pulse2percept_2017}.

\subsection{Model Fitting}
The open parameters of each model were learned to minimize the mean-squared error (MSE) between predicted phosphene brightness and ground truth \cite{perez_fornos_temporal_2012} at each time step using the Powell and Nelder-Mead methods provided by SciPy's \texttt{minimize} function \cite{virtanen_scipy_2020}.
To test a model's capacity to \textit{describe} the data, we fit each individual panel in the ground-truth data \cite{perez_fornos_temporal_2012}.
To test a model's ability to \textit{predict} unseen data, we performed leave-one-subject-out and leave-one-stimulus-out cross-validation, where the model was trained on data from all but one subject or stimulus condition, respectively, and then tested on the held-out data.
Performance was quantified using MSE and Pearson's correlation coefficient ($r$).

All our code is available at \url{https://github.com/bionicvisionlab/2024-Temporal-Model}.

\section{Results}

Fig.~\ref{fig:results-model-fits} demonstrates the \textit{descriptive} capacities of the spectral ($m=4$) and exponential models, compared to baseline \cite{avraham_retinal_2021}.
Here, each model was individually fit to ten representative time courses of phosphene brightness \cite{perez_fornos_temporal_2012}.
Panels A--C represent prototypical examples where brightness continuously fades with repeated stimulation (stimulus duration is indicated with a gray bar on the top of each panel) and then quickly returns to zero after stimulus offset.
Sometimes, phosphene brightness plateaus at a nonzero value over time (Panels E and H), or persists past the stimulus duration (Panels F and G).
Other times, implant users experience another bright flash of light upon stimulus \emph{off}set (Panels D and H).
As is evident from the figure, both models have the capability to fit these complex time courses, whereas the baseline model does not.


Table~\ref{tab:results-per-subject} reports the \textit{predictive} capacities of our models using leave-one-subject-out cross-validation.
As shown in Table I, for the cross-stimulus procedure, the spectral model outperformed the exponential and the baseline models for six out of nine subjects, achieving Pearson correlation coefficients as high as $0.880$.
For Subjects 4 and 6, the exponential model scored slightly better than the spectral model.
All three models struggled to generalize to the data reported by Subject~2.

When predicting data from unseen stimulus conditions (Table~\ref{tab:results-per-stimulus}), the spectral and exponential models achieved similar scores, again consistently outperforming the baseline model \cite{avraham_retinal_2021} across all stimulus conditions.

\begin{figure}[t!]
    \centering
    \includegraphics[width=\columnwidth]{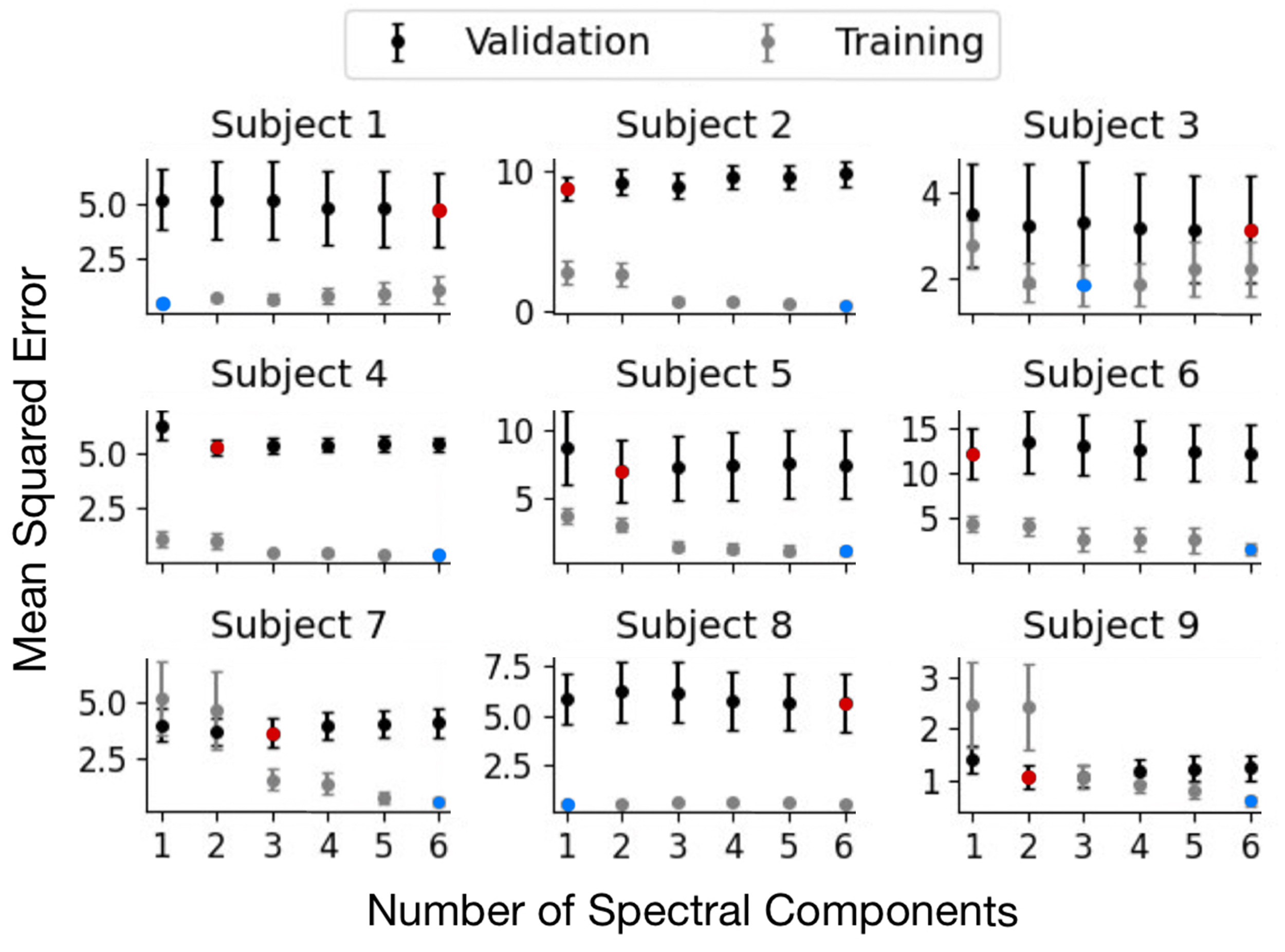}
    \caption{
    Training (gray) and validation (black) MSE $\pm$ Standard Error of the spectral model as a function of the number of spectral components ($m$), presented for each subject and averaged across stimulus conditions. The minimum training MSE is highlighted in blue, and validation in red.
    }
    \label{fig:results-fft-components}
\end{figure}

We further investigated how the number of spectral components ($m$) affects the spectral model's ability to generalize to new data (Fig.~\ref{fig:results-fft-components}).
As some brightness time courses seemingly consisted of multiple decaying rates and a second brightness spike upon stimulus offset, we expected model performance to increase with $m$.
Although this was true for the error on the training set (red curve in Fig.~\ref{fig:results-fft-components}), we found that the smallest generalization error (blue curve) was typically achieved with just one or two spectral components, minimizing overfitting.

\section{Discussion}
We introduced two computational models that are able to capture the complex temporal dynamics of phosphene perception reported by retinal implant users.
Although other work has modeled phosphene fading and persistence \cite{avraham_retinal_2021}, they have yet to be evaluated on real patient data.
Here we demonstrate that the spectral model, which expresses fading and persistence as a truncated Fourier series, is able to generalize to held-out data from different subjects and stimulus conditions.

An additional benefit of the spectral model is that the number of spectral components ($m$) can be selected to capture brightness time courses of differing complexity.
Despite this flexibility, we found that two components are typically sufficient to predict the data.
This makes the model not just computationally efficient, but consistent with previous work highlighting multiple time courses of ganglion cell desensitization in response to repeated electrical stimulation \cite{freeman_multiple_2011, cai_response_2011}.

Despite these promising results, data availability remains a bottleneck for computational work in the field of prosthetic vision.
Given that the reported time courses varied drastically across subjects and stimulus conditions, the predictive capabilities of the spectral and exponential models (especially on Subject~9) are commendable.
Nevertheless, it is interesting to note that all three models struggled to explain Subject~2's data.

In the future, we hope that more data of the kind collected by Perez-Fornos and colleagues \cite{perez_fornos_temporal_2012} can inform the development of predictive models for prosthetic vision.
As stimulus optimization remains an active area of research \cite{granley_hybrid_2022,de_ruyter_van_steveninck_end--end_2022,granley_human---loop_2023}, it is imperative that these algorithms can be trained on accurate and behaviorally validated phosphene models.



\bibliographystyle{IEEEtran}
\bibliography{2024-EMBC-Temporal.bib}
\end{document}